\documentclass[twocolumn,showkeys,preprintnumbers,amsmath,amssymb, prl]{revtex4}

\usepackage{stmaryrd}
\usepackage{txfonts}
\usepackage{amssymb}
\usepackage{mathrsfs}
\usepackage{graphicx}
\usepackage{dcolumn}
\usepackage{bm}
\usepackage{epsfig}

\begin{document}

\title{Radiation of Terahertz Electromagnetic Waves from Build-in Nano Josephson Junctions of Cuprate High-$T_c$ Superconductors}

\author{Shi-Zeng Lin and Xiao Hu}

\affiliation{\(^{1}\)WPI Center for Materials
Nanoarchitectonics, National Institute for Materials Science, Tsukuba 305-0044, Japan\\
\(^{2}\)Japan Science and Technology Agency, 4-1-8 Honcho,
Kawaguchi, Saitama 332-0012, Japan}

\date{\today}

\begin{abstract}
The nano-scale intrinsic Josephson junctions in highly anisotropic cuprate superconductors have potential for generation of terahertz electromagnetic waves. When the thickness of a superconductor sample is much smaller than the wavelength of electromagnetic waves in vacuum, the superconductor renders itself as a cavity. Unlike conventional lasers, the presence of the cavity does not guarantee a coherent emission because of the internal degree of freedom of the superconductivity phase in long junctions. We study the excitation of terahertz wave by solitons in a stack of intrinsic Josephson junctions, especially for relatively short junctions. Coherent emission requires a rectangular configuration of solitons. However such a configuration is unstable against weak fluctuations, contrarily solitons favor a triangular lattice corresponding to an out-phase oscillation of electromagnetic waves. To utilize the cavity, we propose to use an array of stacks of short intrinsic Josephson junctions to generate powerful terahertz electromagnetic waves. The cavity synchronizes the plasma oscillation in different stacks and the emission intensity is predicted to be proportional to the number of stacks squared.

\end{abstract}


\maketitle

\section{Introduction}

Generations of electromagnetic (EM) waves are mainly based on two principles; the low frequency waves are generated by electronic oscillations while the high frequency waves are generated by photonic methods. Because of the fundamental electron-velocity limits, the performance of microwave electronic circuits degrades very rapidly above 100GHz. The photonic generators are also limited to the frequency higher than several terahertz (THz). There exists a frequency gap ranging roughly from 0.1THz to 10THz, which is usually dubbed as the THz gap. The THz waves, on the other hand, have potential for wide applications, such as drug detection, security check and so on\cite{Ferguson02,Tonouchi07}. Thus it is extremely demanding to develop solid-state, compact generators of THz EM waves. The Josephson effect in cuprate high-$T_c$ superconductors is promising for this purpose\cite{HuReview09}.

When two superconductors get sufficiently close to each other to form a tunnel junction or Josephson junction, macroscopic quantum phase coherence will be established between them. According to the uncertainty relation between the number of Cooper pairs and the superconductivity phase, a supercurrent appears in the tunnel junction, which is given by the dc Josephson relation\cite{BaroneBook}
\begin{equation}\label{eq1}
I  = I_c \sin P,
\end{equation}
with $P$ being the phase difference between these two superconductors and $I_c$ the critical current.
Biased by a dc voltage $V$, the Josephson junction is essentially a two-level system with energy difference $2eV$. As Cooper
pairs are tunneling, EM wave is emitted from the junction, whose frequency is governed by the ac Josephson relation
\begin{equation}\label{eq1}
\hbar \omega=\hbar \partial_t P=2eV.
\end{equation}
For instance, $1$mV corresponds to $0.483$THz. In principle, the operating frequency can be tuned continuously by the biased voltage. In contrast to conventional lasers, the population inversion can be achieved in a meta-stable superconducting state with finite normal resistance realized in a junction with small conductance. The radiation from a single junction, however, is only about 1pW and the frequency is below THz limited by the small superconducting energy gap\cite{Yanson65,Langenberg65}.

\begin{figure}[t]
\psfig{figure=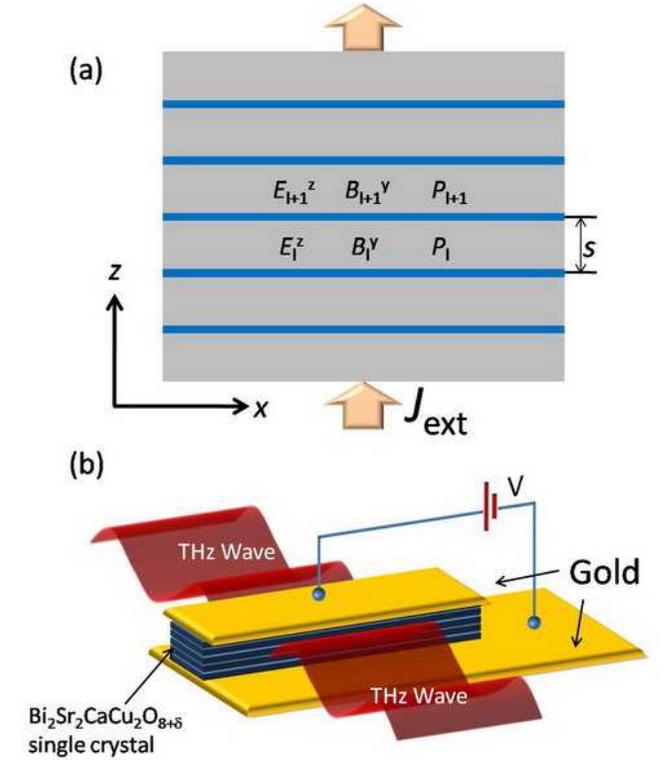,width=\columnwidth} \caption{\label{f1}(a) Highly anisotropic cuprate superconductors like BSCCO can be modeled as a stack of IJJs. The blue (gray) region denotes superconducting (insulating) layers. (b) Proposed device to generate THz EM waves. The BSCCO single crystal is sandwiched by two gold electrodes through which a dc current is fed in.}
\end{figure}

One may fabricate an array of Josephson junctions to enhance the radiation power\cite{Jain84,Barbara99}. Actually, it was found in 1992 that layered high-$T_c$ superconductors, such as $\rm{Bi_2Sr_2CaCu_2O_{8+\delta}}$(BSCCO), intrinsically form a stack of Josephson junctions (IJJs) of nano-scale thickness\cite{Kleiner92}, as drawn schematically in Fig. \ref{f1}(a). The advantages of IJJs are: 1) large superconductivity energy gap, which allows for operations up to 15THz; 2) almost homogeneous in the atomic scale, which makes superradiation possible; 3) these nano-scale build-in junctions render a huge inductive coupling between neighboring junctions. The proposed device using BSCCO single crystal to generate THz wave is depicted in Fig. \ref{f1}(b).

The following two conditions must be met in order to achieve strong THz emission. First, coherent emission of EM waves involves synchronized oscillations of individual elements. Synchronization requires mutual couplings between each oscillator and usually involves long-range interactions. In the context of the Kuramoto model\cite{Acebron05}, it was found that the lower critical dimensions for long-range synchronous oscillations is four\cite{Hong05}. Practically, one introduces a common resonator that couples to all oscillators. This resonator provides a feedback mechanism and establishes a long-range interaction between oscillators, which forces all oscillators to run at its resonant frequency. For one dimensional array of Josephson junctions connected to a common RLC resonant circuit, it was demonstrated rigorously that the system can be mapped to the infinite dimensional Kuramoto model\cite{Wiesenfeld96}, where each junction is coupled to the rest junctions. For a stack of IJJs, it was formulated theoretically\cite{Bulaevskii06PRL,Koshelev08} that there exists a significant impedance mismatch for a stack of thickness of several micrometers as in the recent experiments\cite{Ozyuzer07,kadowaki08}, similar to that in a thin capacitor. Thus a thin stack of IJJs intrinsically forms a cavity, which is helpful for synchronizing all junctions.

Secondly, for a stack of IJJs to work as an efficient gain medium, the supercurrent should be modulated in the lateral directions such that considerable dc input power can be pumped into the Josephson plasma oscillation, a collective composite wave of electromagnetic oscillations and back-and-forth tunneling of Cooper pairs. One well-known example of the modulation in a single Josephson junction is $2\pi$ change in the superconductivity phase, corresponding to a shuttling soliton, a nonlinear quantized topological excitation. Another example is the recently discovered new dynamic state in a stack of strong inductively coupled IJJs\cite{szlin08b,Koshelev08b,Hu08,szlin09a,Hu09}, where $\pi$ phase changes are localized at the nodes of electric fields and stacked periodically along the $c$ axis. By exciting the in-phase plasma mode, this $\pi$ kink state supports strong THz emissions.

Yet a cavity becomes a standard ingredient in most lasers and is almost indispensable to coherent radiation. However, for a stack of long IJJs embedded in the cavity formed by the stack itself, because of the existence of internal degree of freedom, the superconductivity phases of different junctions may organize themselves in a way that the oscillations in neighboring junctions are out-phase. To be concrete, let us consider the stack shown in Fig. \ref{f1}(a). The stack forms a two dimensional cavity and the plasma mode is $\cos(m\pi x/L)\sin(q\pi z/L_z)$ or is simply denoted as $(m, q)$ mode. Here $L$ is the length of the stack and $L_z$ is the thickness. Coherent emission from edges at $x=0$ or $x=L$ requires the excitation of $q=1$ mode which is not realized in some specific configurations of the superconductivity phase such as solitons as will be discussed later. Therefore it is of both theoretical and practical importance to understand the behavior of superconductivity phase in a resonant cavity.

In the present paper, we investigate the behaviors of solitons in a stack of IJJs, especially in a relatively short one. We find that in a short stack with radiating boundary condition, the soliton becomes incomplete and resembles oscillating standing waves. We also find that the rectangular lattice of solitons associated with in-phase plasma oscillation is unstable. The out-phase oscillation of soliton in a cavity leads us to consider an array of short junctions enclosed in a cavity to avoid the internal structure of phases in individual stacks. This configuration provides an alternative scheme to achieve strong THz radiation from Josephson junctions.

\section{Model}

We assume that the stack is infinitely long in the $y$ direction and the system is reduced into two dimensions as shown in Fig. \ref{f1}(a). The dynamics in a stack of IJJs is governed by the well-known inductively coupled sine-Gordon equations\cite{Sakai93,Bulaevskii96,Koyama96,Koshelev01}
\begin{equation}\label{eq3}
\partial _x^2 P_l  = (1 - \zeta \Delta_{\rm{d}})[\sin P_l  + \beta \partial _t P_l  + \partial _t^2 P_l  - J_{{\rm{ext}}}],
\end{equation}
where $P_l$ is the gauge invariant phase difference at the $l$-th junction, $\zeta$ the inductive coupling of order of $10^5$, $\beta$ the normalized conductivity and $J_{{\rm{ext}}}$ the external current\cite{szlin09a}. Length is in unit of the London penetration depth  $\lambda_c\approx200\rm{\mu m}$ and time in unit of the Josephson plasma frequency $\omega_J\approx0.5\rm{THz}$. $\Delta_{\rm{d}}$ is the finite difference operator defined as $\Delta_{\rm{d}} f_l\equiv f_{l+1}+f_{l-1}-2f_l$. We use $\beta=0.02$ and $\zeta=7.1\times 10^4$, which are typical for BSCCO.\cite{szlin09a} The normalized magnetic field $B_l$ and electric field $E_l$ in $l$-th junction are related to $P_l$ by $(1 - \zeta \Delta_{\rm{d}})B_l^y=\partial_x P_l$ and $E_l=\partial_t P_l$. Because of the significant impedance mismatch, the oscillating magnetic field is extremely small at the edges, thus as a good approximation, the boundary condition can be written as
\begin{equation}\label{eq4}
\partial _t P_l=Z\partial _x P_l,
\end{equation}
with the impedance of the stack $Z\sim\lambda_{\rm{EM}}/L_z\gg1$. Here $\lambda_{\rm{EM}}\sim300\rm{\mu m}$ is the wavelength of THz EM wave in vacuum and $L_z\sim1\rm{\mu m}$ is the thickness of the stack. We consider a situation where no external magnetic field is applied to the stack. Equation (\ref{eq3}) together with the boundary condition Eq. (\ref{eq4}) is solved numerically and the electromagnetic fields are then derived from $P_l$.

\begin{figure}[t]
\psfig{figure=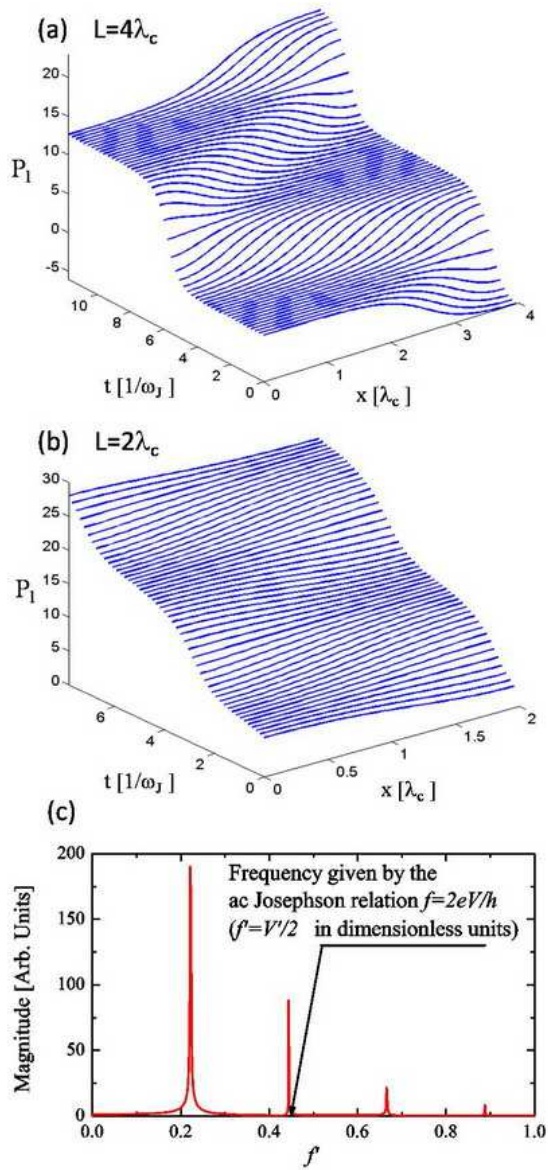,width=\columnwidth} \caption{\label{f2}(a) Time evolution of $P_l$ in a long junction $L=4\lambda_c$ and with the closed boundary condition $Z=\infty$. (b) Time evolution of $P_l$ in a relatively short junction $L=2\lambda_c$ and with radiating boundary condition $Z=10$. The soliton under this situation becomes incomplete. (c)  Frequency spectrum of the electric field at a voltage of $V=2.79$ associated with the first zero-field step and $J_{\rm{ext}}=0.1$ for $L=2\lambda_c$. The fundamental peak does not come to the frequency predicted by the ac Josephson relation from the voltage.}
\end{figure}

\section{Solitons solution and its stability}
Equation (\ref{eq3}) exhibits rich dynamical behaviors\cite{HuReview09} and one possible dynamic state is the solitons state. A soliton carries quantum flux $\Phi_0=hc/2e\approx2.0\times10^{-15}\rm{Wb}$, and it experiences repulsive force from other solitons with the same polarity. The interplay between the interaction between solitons and the cavity gives rise to rich configurations of solitons, from the rectangular lattice to triangular one. For a rectangular configuration of solitons along the $z$ axis, Eq. (\ref{eq3}) is decoupled and the problem essentially becomes single-junction physics. We first discuss the properties of the uniform soliton solution and then consider its stability.

A soliton has characteristic length of $\lambda_c$ and undergoes Lorentz contraction when its velocity approaches the plasma velocity, i.e. the light velocity in junctions. Thus, to support a pronounced soliton structure, the length of junction should be much larger than $\lambda_c$. As shown in Fig. \ref{f2}(a) for $L=4\lambda_c$, a well-defined soliton propagates leftwards and get reflected at the left edge and becomes an anti-soliton. This anti-soliton then travels rightwards and repeats the above process. In one period, the phase increases by $2\pi$ and the period is given by the voltage according to Eq. (\ref{eq1}). The periodic motion and reflection of solitons inside the cavity excite plasma oscillation at a frequency depending on the pattern of solitons in the $x$ direction, and induce current steps at voltages $V=2n\pi/L$ with $n$ being the number of solitons in the lateral direction \cite{szlin09a}.

It would be interesting to study the structure of soliton in a junction with length comparable to $\lambda_c$ and with slightly open boundary condition. In this case, it is difficult to see a complete soliton traveling across the junction and the dynamics of $P_l$ resemble oscillations of standing waves in addition to a linear term in time, as shown in Fig. \ref{f2}(b). However, because of the commonly underlying parametric instability for solitons, the frequency spectrum is the same as that of long junction, namely many frequency harmonics are visible, and the radiation frequency and voltage does not obey the ac Josephson relation as depicted in Fig. \ref{f2}(c), opposed to the recent experiments\cite{Ozyuzer07}.

In a recent work, a new state exhibiting standing wave oscillations but with the same \emph{IV} characteristics as those of solitons solution was found in a relatively short junction\cite{Tachiki09}. Our simulations of solitons in a junction with length comparable to $\lambda_c$ reproduce all the results in Ref. \cite{Tachiki09}, which lead one to assign the dynamic state in Ref. \cite{Tachiki09} as a soliton state.

We then investigate the stability of the uniform soliton solution by including small fluctuations into the system, e.g. adding white-noise currents at the right-hand side of Eq. (\ref{eq3}). The uniform solution then becomes unstable and the system evolves into the $\pi$ kink state. The uniform solution cannot be stabilized by introducing dissipation, radiation or a small magnetic field either. Indeed, due to the strong mutual repulsive interaction, the solitons favor triangular lattice corresponding to the $q=N$ cavity mode with $N$ the number of junctions in the stack. Thus in the soliton state, due to the internal degree of freedom, the cavity does not guarantee a coherent oscillations in all junctions.

\section{An Array of Stacks of IJJs}
\begin{figure}[t]
\psfig{figure=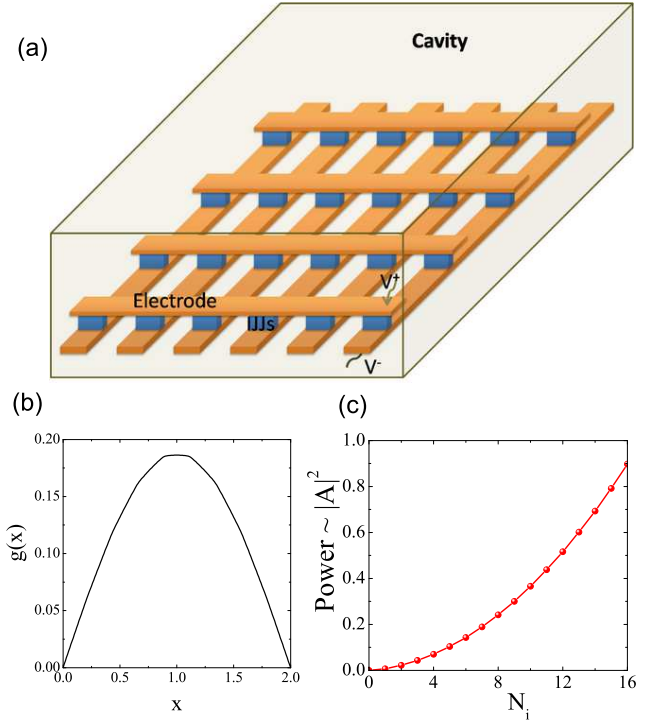,width=\columnwidth} \caption{\label{f3}(a) Schematic view of an array of IJJs embedded in an external cavity. (b) Profile of $g(x)$ for the first cavity mode. (c) Dependence of the radiation power $S_r\sim|\omega A|^2/(2Z_c)$ on the number of IJJs stacks $N_i$. In obtaining the results in (b) and (c), we assume that IJJs stacks with length $10\rm{\mu m}$ are evenly distributed in the cavity with dielectric constant $\epsilon_d=0.1$. Here $L_c=400\rm{\mu m}$, and $N_i=8$ in (b).}
\end{figure}

To utilize the cavity, one way is to adopt an array of stacks of short IJJs to avoid the complicated behaviors of $P_l$ in individual stacks. A schematic view of such a setup is depicted in Fig. \ref{f3}(a).

We consider short IJJs with $L\ll\lambda_{\rm{c}}$, thus the soliton state cannot be stabilized. The short IJJs also allow
for efficient cooling and less Joule heating\cite{Krasnov01}. For simplicity, the height of the cavity is assumed to the same as of IJJs and $L_y$ is assumed to be infinite as before. For $L_z\ll\lambda_{\rm{EM}}$, only the mode with the wave vector $k_z=0$ can propagate in vacuum. This uniform mode tends to excite in-phase mode in IJJs. The electromagnetic
waves in the cavity are therefore governed by
\begin{equation}\label{eqa1}
\partial _x^2P=J_c'(x)\sin  P+\beta '(x)\partial _tP+\epsilon '(x)\partial _t^2P-J_{\text{ext}}'(x).
\end{equation}
In IJJs, Eq. (\ref{eqa1}) is the uniform-mode correspondence of Eq. (\ref{eq3}), while in vacuum $P$ should be interpreted as an auxiliary variable determined by $\partial_x P=B_y$ and $\partial_t P=E_z$. In junctions $\beta '(x)=\beta$, $J_c'(x)=J_c$, $J_{\text{ext}}'(x)=J_{\text{ext}}$ and $\epsilon '(x)=1$, while in vacuum $\beta '(x)$, $J_c'(x)$, $J_{\text{ext}}'(x)$ are zero and $\epsilon '(x)=\epsilon_d$ with $\epsilon_d$ being the dielectric constant of vacuum. We assume a metallic cavity, and at the edges of the cavity $x=0$ or $x=L_c$, we have $E_z=0$ for an ideal cavity. The radiation effect is taken into account by introducing an effective surface impedance $Z_c$ for the cavity. Then the radiation power is given by $S_r=|E_z^2|/(2Z_c)$.

The solution to Eq. (\ref{eqa1}) in the linear region is
\begin{equation}\label{eqa2}
P = \omega t + {\mathop{\rm Re}\nolimits} [ - i A g(\omega_c, x)\exp (i\omega t)],
\end{equation}
where the first term is the rotating phase responsible for the dc voltage and the second term is the uniform plasma oscillations along the $z$ axis. $g(\omega_c, x)$ is the cavity mode with the resonant frequency $\omega_c$, which can be determined by neglecting the Josephson current, external current and dissipations, since the Josephson current and external current are responsible for the excitation of EM waves while dissipation has the effect of broadening the resonance line width. The profile of $g(x)$ for the first mode is shown in Fig. \ref{f3}(b), where there is a half wavelength residing in the cavity. The amplitude of the plasma oscillation $A$ then is given by
\begin{equation}\label{eqa3}
A=\frac{\int_0^{L_c} dx J_c'(x)g(x)}{\int_0^{L_c} dx \left[\left(\omega ^2-\omega _c^2\right)\epsilon '(x)-i \beta '(x)\omega
\right]g(x)^2}.
\end{equation}
The radiation slightly modulates the supercurrent in each stack and pumps energy into the plasma oscillation. The amplification of the plasma oscillation is most efficient for the first cavity mode because the radiation in the region with $g(x)<0$ is anti-phase for higher modes. For the first mode, we have $A\sim N_i$ with $N_i$ being the number of IJJs stacks. Thus the total radiation power increases as $N_i$ squared as shown in Fig. \ref{f3}(c), which presumes a superradiation. It is remarked that IJJs are not necessarily identical since the cavity helps to synchronize them.\cite{Jain84,Barbara99}

\section{Conclusions}
To summarize, we have discussed the possible radiation of terahertz electromagnetic waves from cuprate high-$T_c$ superconductors and the associated dynamics of superconductivity phase. We studied the behaviors of a soliton in a relatively short junction, and found that the soliton becomes incomplete and resembles the oscillating standing waves. We also found the in-phase configuration of solitons is unstable even though they are embedded in a cavity. The instability is caused by the mutual interaction between solitons. Based on these observations, we proposed to integrate an array of stacks of short intrinsic Josephson junctions in an external cavity to attain the powerful terahertz emission. The short junctions avoid complicated behaviors of the superconductivity phase and allow for efficient synchronization by the external cavity. Once they are synchronized, the radiation power increases as the number of stacks squared according to our calculations.

\vspace{3mm}
\noindent{\it Acknowledgement --}
This work was supported by WPI Initiative on Materials Nanoarchitronics, MEXT, Japan and CREST-JST Japan.


\end{document}